\documentclass[apjl]{emulateapj}
\usepackage{color}

\usepackage{graphicx}

\begin{document}



\title{Magellan Adaptive Optics Imaging of PDS 70: Measuring the Mass Accretion Rate of a Young Giant Planet within a Gapped Disk}
\color{black}
\shorttitle{The Mass Accretion Rate of PDS 70b}
\shortauthors{Wagner, Follette, Close, Apai et al.}
\author{Kevin Wagner\altaffilmark{1,2,3,}$^{\star}$, Katherine B. Follette\altaffilmark{4}, Laird M. Close\altaffilmark{1}, D\'aniel Apai\altaffilmark{1,3,5,6}, Aidan Gibbs\altaffilmark{1}, Miriam Keppler\altaffilmark{6}, Andr\'e M\"uller\altaffilmark{6}, Thomas Henning\altaffilmark{6}, Markus Kasper\altaffilmark{7}, Ya-Lin Wu\altaffilmark{8}, Joseph Long\altaffilmark{1}, Jared Males\altaffilmark{1}, Katie Morzinski\altaffilmark{1}, and Melissa McClure\altaffilmark{9}}


\altaffiltext{1}{Steward Observatory, University of Arizona}

\altaffiltext{2}{National Science Foundation Graduate Research Fellow}
\altaffiltext{3}{NASA NExSS \textit{Earths in Other Solar Systems} Team}
\altaffiltext{4}{Department of of Physics and Astronomy, Amherst College}
\altaffiltext{5}{Lunar and Planetary Laboratory, University of Arizona}
\altaffiltext{6}{Max Planck Institute for Astronomy, Heidelberg, Germany}
\altaffiltext{7}{European Southern Observatory, Garching, Germany}
\altaffiltext{8}{Department of Astronomy, University of Texas, Austin}
\altaffiltext{9}{University of Amsterdam, Netherlands}

\altaffiltext{$\star$}{Correspondence to: kwagner@as.arizona.edu}


\begin{abstract}

PDS 70b is a recently discovered and directly imaged exoplanet within the wide ($\gtrsim$40 au) cavity around PDS 70 (\citealt{Keppler2018}, \citealt{Muller2018}). Ongoing accretion onto the central star suggests that accretion onto PDS 70b may also be ongoing.  We present the first high contrast images at H$\alpha$ (656 nm) and nearby continuum (643 nm) of PDS 70 utilizing the MagAO system. The combination of these filters allows for the accretion rate of the young planet to be inferred, as hot infalling hydrogen gas will emit strongly at H$\alpha$ over the optical continuum. We detected a source in H$\alpha$ at the position of PDS 70b on two sequential nights in May 2018, for which we establish a false positive probability of $<$0.1\%. We conclude that PDS 70b is a young, actively accreting planet. We utilize the H$\alpha$ line luminosity to derive a mass accretion rate of $\dot M= 10^{-8\pm1}$ M$_{Jup}/yr$, where the large uncertainty is primarily due to the unknown amount of optical extinction from the circumstellar and circumplanetary disks. PDS 70b represents the second case of an accreting planet interior to a disk gap, and is among the early examples of a planet observed during its formation. 

\end{abstract}


\section{Introduction}

While gapped (``transition" and ``pre-transition") disks around young stars are fundamental to inform planet formation and disk evolution models (e.g., \citealt{DAngelo2003}, \citealt{Kley2012}, \citealt{Uribe2013}), so far the link to planet formation has lacked significant direct evidence. With rapidly advancing instrumentation (e.g., \citealt{Macintosh2006}, \citealt{Close2008}, \citealt{Beuzit2008}) it is now possible to place (often powerful) constraints on massive planets interior to gapped disks around nearby young stars. 



At young ages ($\lesssim$100 Myr), giant planets are hot and luminous enough for their continuum thermal emission to be detectable in the infrared (e.g., \citealt{Mordasini2017}). Indeed, on the order of a dozen super-Jupiters have been discovered in recent years orbiting nearby young stars (see the recent review by \citealt{Bowler2016}). Meanwhile, at very young ages ($\lesssim$10 Myr), giant planets may still be accreting gas from the local disk environment. During active accretion epochs, the shocked hot ($\sim$10,000 K) infalling Hydrogen gas may generate a significant and observable H$\alpha$ luminosity (\citealt{Zhu2015, Eisner2015}) that can easily boost planet-to-star contrast ratios to higher levels than in the infrared. Furthermore, a detection at H$\alpha$ also enables a mass accretion rate to be derived from empirical accretion rate vs. line luminosity relations (e.g., \citealt{Rigliaco2012}), thereby enabling planet formation to be observed as a time dependent process. This is the motivation behind the high-contrast H$\alpha$ capabilities of the Magellan Adaptive Optics System (\citealt{Close2012}, \citealt{Morzinski2016}) on the 6.5-m Magellan Clay Telescope, its flagship Giant Accreting Protoplanet Survey (GAPlanetS: Follette et al., \textit{submitted}), and the H$\alpha$ capabilities of the Very Large Telescope's Spectro-Polarimetric High Contrast Exoplanet Research Experiment (VLT/SPHERE). The power of this approach has already been demonstrated with MagAO's detection of H$\alpha$ from the accreting protoplanet LkCa 15b \citep{Sallum2015}. 

PDS 70 is a K7 pre-main sequence T Tauri type star in the Upper Centuarus Lupus association (\citealt{Riaud2006}, \citealt{Pecaut2016}). The star hosts a gapped disk of moderate inclination (i$\sim$50$^\circ$), with the gap extending from $\lesssim$17 au to 60 au. This region ($\sim$0$\farcs$2 to 0$\farcs$6) is directly accessible to the high-contrast search zones of current adaptive optics (AO) systems (\citealt{Hashimoto2012}, \citealt{Dong2012}, \citealt{Long2018}), providing motivation for direct imaging searches to test the hypothesis that PDS 70's gap has been cleared by the recent formation of one or more giant planets, as generally predicted by \cite{Dodson2011}. 



Indeed, an on-going survey using VLT/SPHERE has recently discovered thermal emission from a giant planet interior to the disk's gap (\citealt{Keppler2018}, \citealt{Muller2018}). The planet is confirmed in multiple photometric bands, with multiple telescopes and instruments, and in multiple epochs. In this configuration, PDS 70b is likely responsible for clearing and maintaining the gap, thereby driving its own mass accretion. A detection of PDS 70b in H$\alpha$ would enable a mass accretion rate to be estimated for the young planet, which has only been performed for one other exoplanet (LkCa 15b: \citealt{Sallum2015}, Follette et al. \textit{submitted}). 




\section{Observations and Data Reduction}

We observed PDS 70 using MagAO's visible (VisAO: \citealt{Males2014}, \citealt{Close2014}) camera on 2017 Feb 10 as part of the GAPlanetS program (PI: Follette) and on two nights of general observing time (PI: Wagner) on 2018 May 3, and 2018 May 4. Each of our three observations was executed in the angular and spectral differential imaging plus mode (SDI+, \citealt{Close2018}). The new mode utilizes a spinning half wave plate in the fore-optics to randomize and effectively equalize any polarized disk signals in either beam, which then cancel along with the diffraction pattern in the spectral differential imaging (SDI) step.




In general, each observing sequence consisted of 1$-$2 hours of dithered observations. The conditions were excellent ($\sim$0$\farcs$5 seeing) and the amount of field rotation was $\sim$90$^\circ$ on each night. The core of PDS 70 was unsaturated in each of our exposures, enabling efficient field centering and frame-by-frame photometric calibration. Despite similar conditions, in 2017 the $R$-$I$-band brightness of PDS 70 measured by the wavefront sensor was 44\% fainter in both H$\alpha$ and continuum filters. The All Sky Automated Survey for SuperNovae \citep{asassn} has established a variability amplitude that is similar to or exceeding 40\% for PDS 70 within the past year, so it is reasonable to assume that the different brightness observed by MagAO in 2017 vs. 2018 may be intrinsic. This decrease in brightness and corresponding decrease in AO performance likely precluded the detection of such a faint companion at H$\alpha$ in Feb 2017. 




The raw data were dark-subtracted, divided by the instrumental flat field, and divided by the detector integration time of 30 seconds (45 sec for Feb 2017). The frames were aligned via cross-correlation with bi-linear interpolation to account for sub-pixel shifts. The position of PDS 70 was found via a Gaussian fit to the median PSF, and a second centering step was then performed utilizing the rotational symmetry of the PSF. This second step resulted in less than half of a pixel correction ($\lesssim$4 mas) compared to the Gaussian fit. The accuracy of this method is estimated to be typically around $\lesssim$0.25 pixels, or $\lesssim$2 mas \citep{Morzinski2015}.  We performed a frame selection based on the counts at the PSF core of PDS 70, and iterated upon this parameter with the presence of injected planets to arrive at the optimal value to maximize the SNR at 0$\farcs$2. This resulted in rejecting frames whose core was less than (100,150,100) counts/s on 2017 Feb 10, 2018 May 3, and 2018 May 4, respectively, which correspond to (23\%, 48\%,43\%) of frames being rejected, and total integration times of (109,68,82) minutes. 

The cubes were PSF subtracted through two independent algorithms: 1) classical angular differential imaging (cADI: \citealt{Marois2006}) and 2) projection onto eigenimages (Karhunen-Lo\`eve Image Processing, or KLIP: \citealt{Soummer2012}) via self-developed IDL scripts \citep{Apai2016}. Prior to cADI, the individual images were high-pass filtered by subtracting a 11x11 pixel (square) median-smoothed version of the image. For KLIP, we modelled and subtracted the PSF in six annular segments in the radial range of 0$\farcs$1-0$\farcs$3 from PDS 70, and iterated upon the remaining parameters in the presence of injected planets (similar to the strategy outlined in \citealt{Meshkat2014}). The parameters that we explored for KLIP included high-pass filter width, minimum and maximum reference angle separation, and number of principal components. We arrived at optimal values of high-pass filter width = (15, 13, 17) pixels, minimum reference angle = (1.1, 3.3, 5.5)$^\circ$, maximum reference angle = (40, 40, 45)$^\circ$, and number of principal components = (5, 4, 4), for the three sequential epochs. 


Following either PSF subtraction, the cubes were derotated and combined with a noise-weighted mean, which assigns lower weights to individual pixels with higher noise in the final derotated and combined image \citep{Bottom2017}. The H$\alpha$ and continuum images were then convolved with the measured width of the PSF (FWHM$\sim$7 pixels). The continuum image was magnified by a factor of 656/643 (2\% to account for radial scaling of the diffraction pattern with wavelength) and then the flux was scaled by the ratio of the peak counts of the median H$\alpha$ PSF to the median continuum PSF. The final SDI+ images were generated by subtracting the scaled continuum image from the H$\alpha$ image.


We also processed the data through a third pipeline, the GAPlanetS pipeline, which is described in detail in Follette et al., \textit{submitted}. This pipeline provides a complete independent reduction from start to finish (notably, the cADI and KLIP pipelines mentioned above share the same initial processing). Briefly, the GAPlanetS pipeline includes dark current subtraction, flat fielding, cosmic ray rejection, rotational symmetry based star-centering, and utilizes the public pyKLIP package for PSF subtraction \citep{Wang2015}. The PSF subtraction is similar to our previously described KLIP pipeline with the exception that there is no maximum rotation angle imposed upon the reference library. Optimization of PyKLIP parameters was done by maximizing the signal to noise ratio of  planets injected into the continuum images. We utilized the first two principal components in the PSF subtraction as this maximized the SNR of the injected planets, and note that the result is not significantly affected by the choice of between 1-50 principal components.  


\begin{figure*}[htpb]
\figurenum{1}
\epsscale{1.1}
\plotone{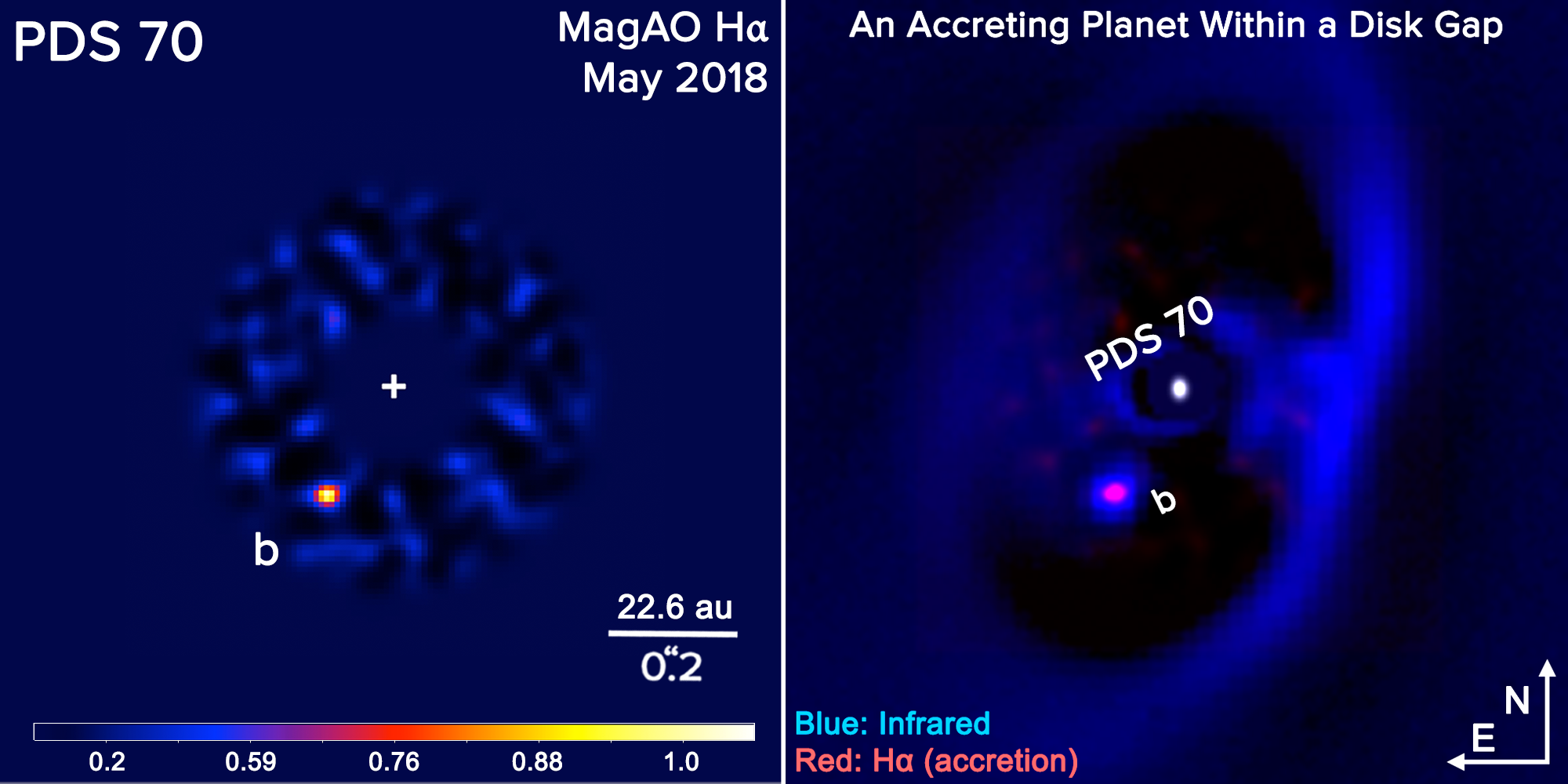}
\caption{Left: MagAO H$\alpha$ SDI+ image of PDS 70. PDS 70b is the only clearly detected point source at $>$95\% confidence. Right: Schematic false color diagram of the components of the PDS 70 system. The image is assembled from the H$\alpha$ image tracing accretion onto the planet (red) and the infrared image (blue) showing the planet's thermal emission and starlight scattered by the disk \citep{Muller2018}. Note that the primary star also has H$\alpha$ emission that is not shown here due to the difficulty in capturing the extreme contrast ratio.}
\end{figure*}


\epsscale{1.1}

\begin{figure*}[htpb]
\figurenum{2}
\epsscale{1.15}
\plotone{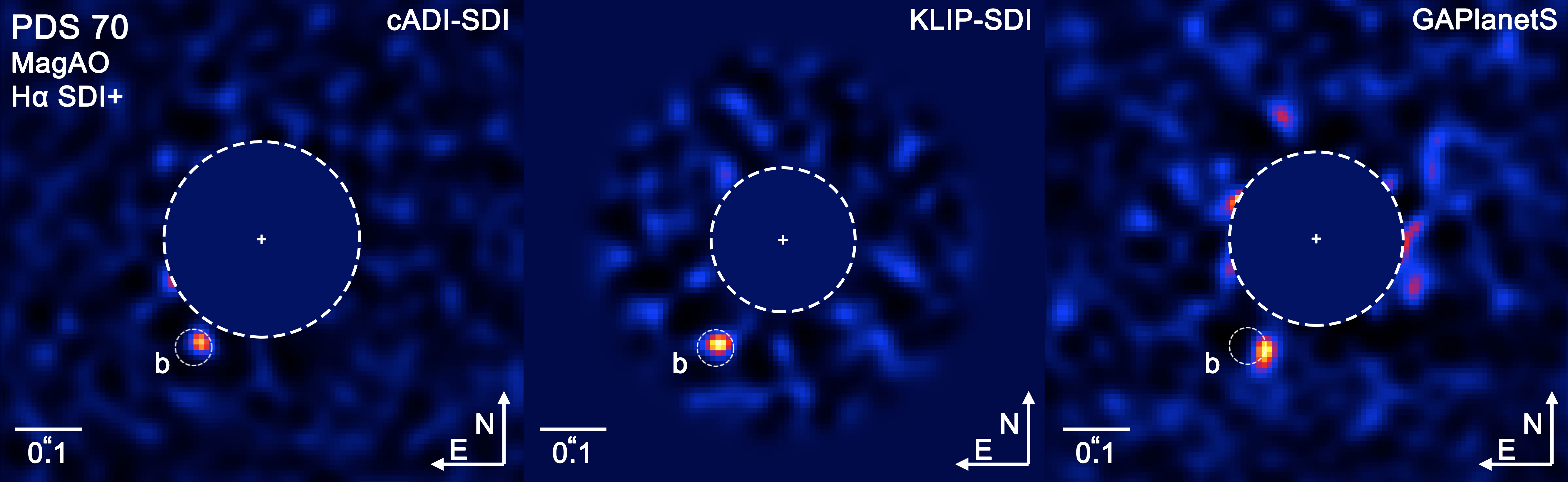}
\caption{2018 May 3/4 MagAO H$\alpha$ SDI+ data reductions utilizing cADI for the angular step (left), utilizing KLIP-ADI (center), and utilizing the GAPlanetS pre-processing and pyKLIP+SDI pipeline (right). All three methods yield a consistent detection of PDS 70b. The smaller dashed white circle (diameter = FWHM) indicates the most recent position of PDS 70b in \cite{Muller2018}, which is consistent with the H$\alpha$ detection's position in each pipeline (see the discussion on astrometric uncertainties in \S3.1). The color scale is normalized and identical to that in the left panel of Fig. 1.}
\end{figure*}

\begin{figure}[htpb]
\figurenum{3}
\epsscale{1.18}
\plotone{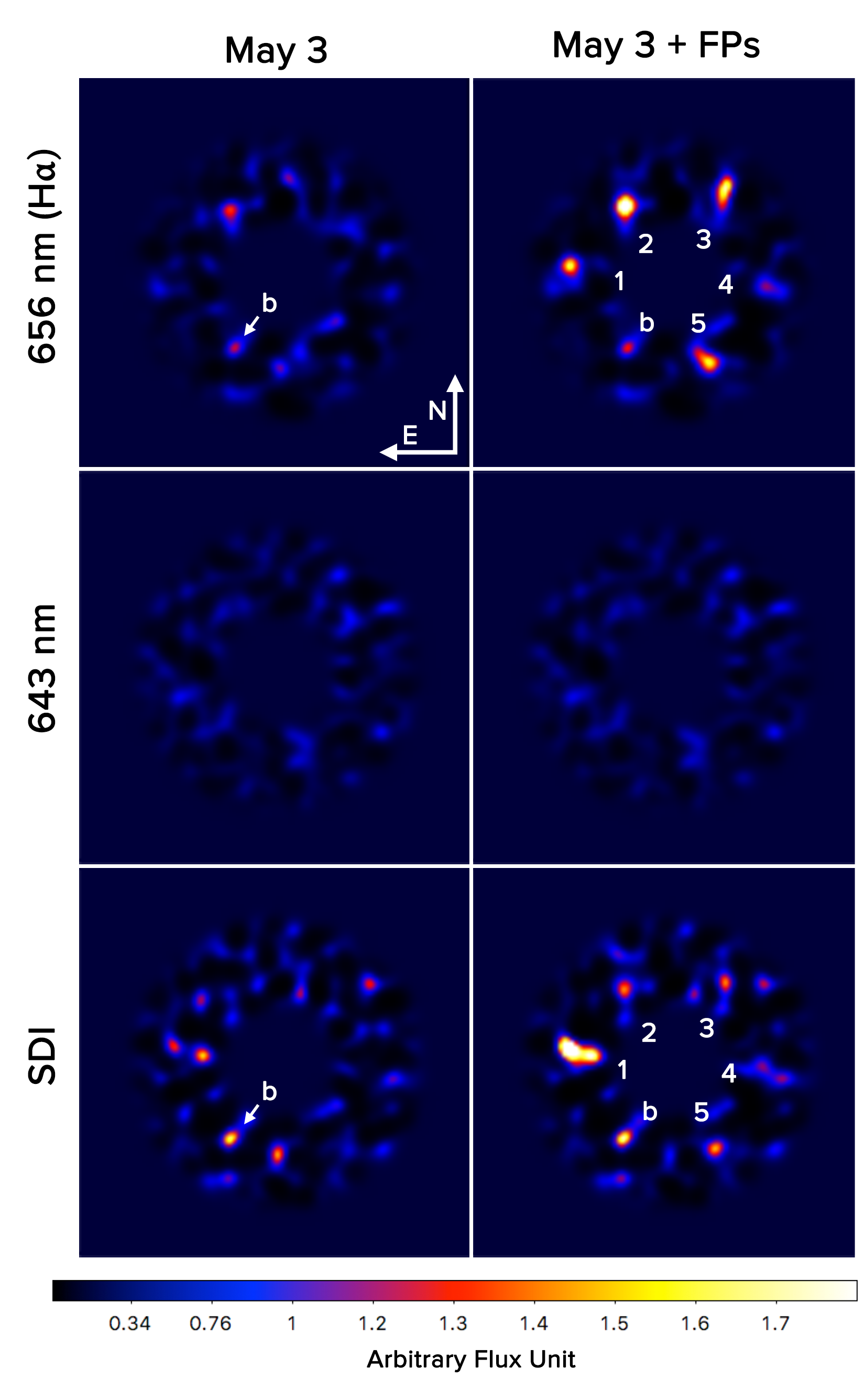}
\caption{2018 May 3 continuum, H$\alpha$, and H$\alpha$ $-$ continuum (SDI) images (left column), and the same data with five injected planets (FPs) at 1.5$\times$10$^{-3}$ contrast (right column).}
\end{figure}

\begin{figure}[htpb]
\figurenum{4}
\epsscale{1.2}
\plotone{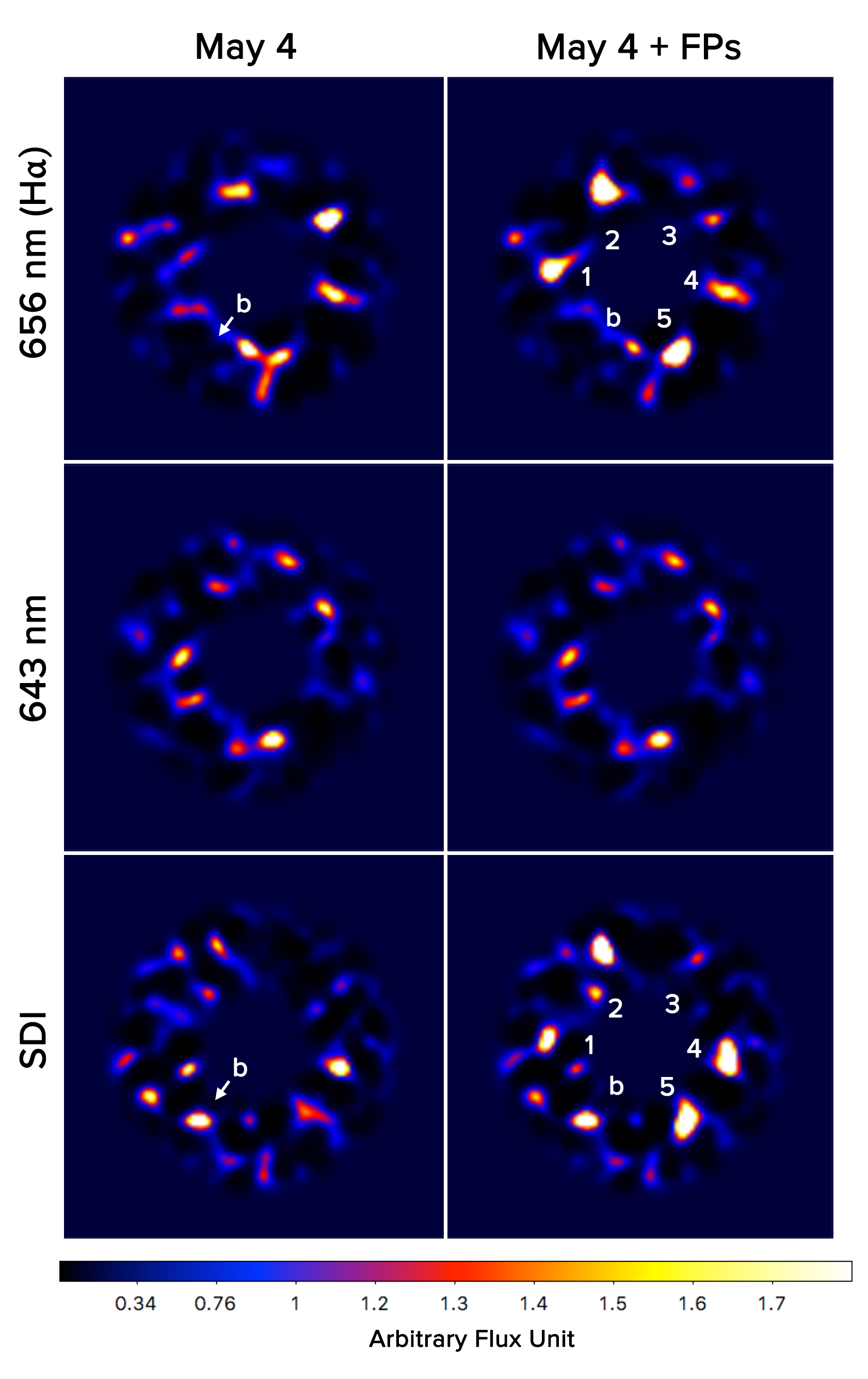}
\caption{2018 May 4 continuum, H$\alpha$, and H$\alpha$ $-$ continuum (SDI) images (left column), and the same data with five injected planets (FPs) at 1.5$\times$10$^{-3}$ contrast (right column).}
\end{figure}

\section{Results}

We detected a tentative ($\sim$2-3$\sigma$) H$\alpha$ point source in the May 3 data following an initial reduction of the data on the morning following the observations. We obtained the position angle and separation of the planet candidate identified by Keppler et al. through private communication and determined that the source identified in H$\alpha$ was in the location of PDS 70b. We observed the source again the following night, and obtained a consistent detection, which combined with the data from the night before yielded a more significant detection ($\sim$4$\sigma$, see Fig. 1). The comparison of the output from our three data reduction pipelines is shown in Fig. 2. The individual channels are shown in Fig. 3 and Fig. 4 for May 3 and 4, respectively. 

While no source is obvious in the single 2017 epoch, it would have been very near to the detection limit (likely due to the faintness of the star in 2017 and weaker AO performance), and we estimate a $\sim$50\% probability that a source of equal brightness would not have been detected. Thus, we consider only the higher quality 2018 data in the proceeding analysis. While all thee pipelines provide consistent results, the following analysis is based on our KLIP+SDI pipeline (\citealt{Apai2016}, center panel Fig. 2) in which PDS 70b is detected at the highest SNR.





\subsection{Astrometry of PDS 70b}


We measured astrometry via centroiding on the source, and established uncertainties through repeating the analysis on injected planets to establish measurement uncertainties. We converted image coordinates to on-sky positions via a platescale calibration of 7.851$\pm$0.015 $mas/pixel$ and true North calibration of $0.9^\circ\pm0.3^\circ$ E of N (\citealt{Close2013}). On May 3 we detected PDS 70b at a separation of 183$\pm$18 mas and position angle (PA) of 148.8$^\circ \pm$1.7$^\circ$. On May 4, we detected PDS 70b at a separation of 193$\pm$12 mas and PA of 143.4$^\circ \pm$4.2$^\circ$. Both of these positions are consistent within 1-$\sigma$ of each other and with the most recent SPHERE astrometry of 192$\pm$8 mas separation and PA of 147$^\circ\pm$2.5$^\circ$ \citep{Muller2018}.


\subsection{H$\alpha$ Luminosity of PDS 70b and the Mass Accretion Rate of a Growing Planet}

On each night and in the combined data we compared the flux in a 5-pixel radius aperture centered on PDS 70b to the mean and standard deviation of identical measurements carried out on the injected planets to measure the H$\alpha$ contrast (and uncertainty) of PDS 70b. On May 3 and 4 we measured contrasts of PDS 70b in H$\alpha$ to the optical continuum of 1.04$\pm$0.70$\times$10$^{-3}$, and 1.40$\pm$0.66$\times$10$^{-3}$, respectively. In the combined image, we measured a contrast of 1.14$\pm$0.47$\times$10$^{-3}$. The contrasts listed above correspond to the brightness ratio of PDS~70b in H$\alpha$ to PDS~70 in the adjacent continuum. This non-standard definition of contrast is chosen for the reason that it eliminates the need to correct for variable accretion onto PDS 70 as well as the star's variable chromospheric activity. While the star is also optically variable, between the two nights we measured less than 5\% variability in both filters, which is substantially smaller than the photometric measurement uncertainties for PDS 70b, and hence no correction for optical variability of PDS 70 is needed.

Following the strategy in \cite{Close2014}, we converted the H$\alpha$ luminosity to a mass accretion rate. Briefly, the calculation follows the conversion from H$\alpha$ luminosity, to accretion luminosity, to accretion rate as outlined in \cite{Rigliaco2012}. The $R$-band apparent brightness of the star is somewhat uncertain ($\sim$0.4 mag) and we adopt here $R$=11.7 \citep{Henden2015}. We also assume a planetary radius equivalent to that of Jupiter, a mass of 5-9 M$_{Jup}$ planet \citep{Keppler2018}, and taking into account our photometric uncertainty, calculate a mass accretion rate of  $\dot{M}_{PDS70b} =  10^{-8.7\pm0.3}$ M$_{Jup}/yr$. To account for a different radius of PDS 70b, the mass accretion rates listed here should simply be multiplied by $R_{PDS70b}/R_{Jupiter}$. For illustrative purposes, we account for up to 3.0 mags of optical extinction from the circumplanetary+circumstellar disks and interstellar medium, as well as the wider mass range found in \cite{Muller2018}, and find a plausible range for the mass accretion rate of  $\dot{M}_{PDS70b} =  10^{-8\pm1}$ M$_{Jup}/yr$ (see Fig. 5). Note that in this example the lower limit is still dominated by the measurement uncertainties, but the substantial amount of extinction significantly raises the upper limit to 10$^{-7}$ M$_{Jup}/yr$. While this choice of extinction is arbitrary, it is not extreme, especially if the circumplanetary disk is viewed at a significant inclination. 



\begin{figure}[htpb]
\figurenum{5}
\epsscale{1.3}
\plotone{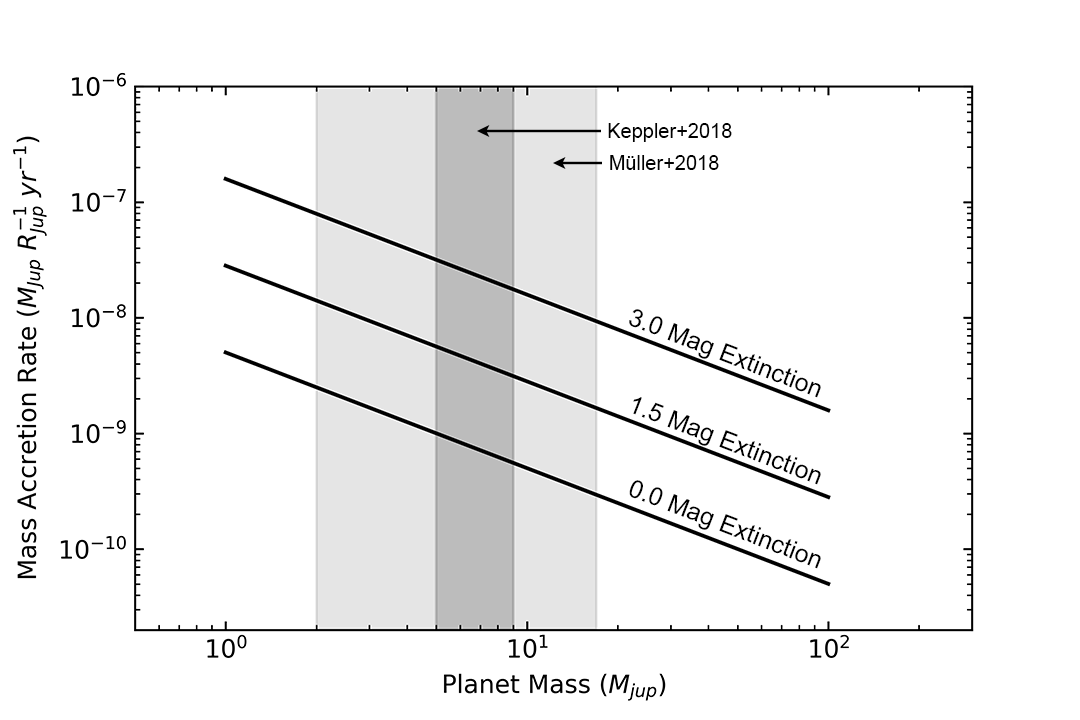}
\caption{The scaling relation of PDS 70b's mass accretion rate with planet mass and extinction. The mass ranges of \cite{Keppler2018} and \cite{Muller2018} are displayed in the shaded regions.}
\end{figure}

\section{Discussion}

\subsection{Probability of a False Positive Result}

To establish the probability of a false positive detection on both nights in 2018, we analyzed the spatial distribution of speckles in the final SDI+ images. Over the image area between 0$\farcs$1$-$0$\farcs$3 from PDS 70, we counted the speckles of similar brightness to the point source at the position of PDS 70b (including the source itself), and found seven such sources for May 3, and twelve for May 4, which are approximately equivalent to 25 and 50 speckles per square arcsecond. These are conservative estimates, as PDS 70b is actually the brightest source in each image and we considered anything of similar brightness to include all sources within a factor of two of its flux to be a plausible false positive.

To be wrongly considered a detection of PDS 70b, a speckle must not only share a consistent brightness, but must also fall within $\sim$2$\sigma$ of the planet's location. We estimated the area corresponding to a 2$\sigma$ astrometric uncertainty through the recovered astrometry of our injected planets, and found this area to be $\sim$0.00043 square arcseconds for May 3 and 0.00072 square arcseconds for May 4. Combined with the speckle densities, these translate into (conservative) false positive probabilities for either night of 1.1\% and 3.6\% for May 3 and 4, respectively, and a combined probability of $\sim$4$\times$10$^{-4}$ for a speckle of the same brightness to appear within 2$\sigma$ of the known location of the object on both nights. Given that the individual detections of PDS 70b fall within 1$\sigma$ of the recent SPHERE astrometry, rather than the 2$\sigma$ criteria considered above, a less conservative estimate may be more appropriate. In this case, the area in question for a speckle to randomly land is reduced by a factor of four, and so is the corresponding false-positive probability.

Confidence in the astrophysical nature of the H$\alpha$ emission from PDS 70b also comes from the significance of the detection in the combined May 3/4 dataset, in which PDS 70b is the only clearly detected source. We estimated the signal to noise ratio by comparing the flux in a 1$\times$FWHM wide aperture compared to the noise measured in all other non-overlapping apertures at the same radius from the star. We followed Equation 9 in \cite{Mawet2014} to estimate the SNR, taking into account the correction for small sample statistics at $\sim$3.5 beam widths from the star. This resulted in 20 independent noise measurements, compared to which the detection of PDS 70b stands out as a 3.9$\sigma$ outlier in the combined 2018 data. Via the same analysis, PDS 70b stands out at 2.6$\sigma$ and 2.4$\sigma$ on May 3 and 4, respectively. As expected for a real source, the SNR improves roughly as the square root of the total exposure time. The corresponding false positive probability is thus $\sim$10$^{-4}$ in the combined image, or around 1\% on either night$-$in excellent agreement with the analysis above. 

\subsection{A Giant Planet Caught in Formation}

The H$\alpha$ emission from PDS 70b indicates that the object is still accumulating mass from its surrounding disk environment. Thus, it is still in the process of formation, and its current mass and mass accretion rate can be used to estimate the final mass that PDS 70b will attain. Considering that PDS 70b has already gained on the order of $\sim$10 M$_{Jup}$ in mass, it is unlikely that PDS 70b will grow even an additional $\sim$10\% of its mass within the disk dispersal lifetime, even at the upper limit of its mass accretion rate of 10$^{-7}$ M$_{Jup}$yr$^{-1}$. Unless the extinction is extremely high, it is unlikely that the accretion rate is higher than the value considered here. In other words, in absence of a dramatic increase in the accretion rate, or extreme cases of optical extinction, it appears that PDS 70b has reached its isolation mass.

We may also make some inferences about the formation history of the young planet by performing the calculation in reverse. Even at its upper limit of 10$^{-7}$ M$_{Jup}$yr$^{-1}$, PDS 70b would have taken at least 20 Myr to grow to its minimum mass estimate of 2 M$_{Jup}$ \citep{Muller2018}. This is four times longer than the system's estimated age \citep{Muller2018}. Thus, to attain its minimum mass in the estimated 5 Myr age of the system, we infer that the accretion rate would have needed to be at least (on average) a modest four times higher than the maximum value considered here.

These inferences are consistent with a formation scenario for PDS 70b in which the planet experienced a period of run-away gas accretion. During this epoch the planet likely (at least partially) played a role in clearing the wide gap that it currently resides in. It is interesting, then, that the planet continues to accrete gas and emit strongly at H$\alpha$. Similar to LkCa 15b \citep{Sallum2015}, the fraction of material accreting onto PDS~70b is comparable to the amount of accretion onto the star ($\dot M_{PDS70}\lesssim10^{-8} M_{Jup}~yr^{-1}$; \citealt{Long2018} and private communication). This suggests that the planet may play a role in shepherding material through the disk gap and onto the star, meanwhile accreting some fraction of this material onto itself. Indeed, \cite{Keppler2018} and \cite{Muller2018} identify several structures extending interior to the gap that may be related to planet-driven spiral waves or other disk transport processes. 

\section{Summary and Conclusions}

We have observed PDS 70b on three nights throughout Feb 2017 to May 2018 with MagAO in the H$\alpha$ SDI+ mode on the 6.5-$m$ Magellan Clay telescope at Las Campanas Observatory, Chile. On sequential nights in 2018, we detected a point source in H$\alpha$ at $\sim10^{-3}$ contrast to the star. Both independent detections in 2018 are at a level of $\sim$2-3$\sigma$, and combined yield a $\sim$4$\sigma$ detection. In 2017, the observations were not sensitive enough to confidently detect the planet.

Both detections in 2018 are consistent within 1$\sigma$ of the other's astrometry, and with the most recent SPHERE astrometric measurement of PDS 70b. We explored the probability that the detection of PDS 70b on both nights is a random false positive through two independent methods, and arrive at a consistent false alarm probability of $\sim$10$^{-4}$. We conclude that the detected H$\alpha$ emission originated from PDS 70b.


We converted the object's H$\alpha$ contrast in 2018 to a mass accretion rate, assuming the range of masses for PDS 70b in \cite{Keppler2018} and \cite{Muller2018}, and several cases of extinction, and found a mass accretion rate of $\dot{M} =  10^{-8\pm1}$ M$_{Jup}/yr$. Given the mass and mass accretion rate of PDS 70b, we estimate that the planet has already acquired $\gtrsim$90\% of its mass. 

\section{Acknowledgments}

The authors express their gratitude toward Korash Assani, Zachary Long, Michael Sitko, and Kaitlin Kratter for useful discussions that improved the quality of this work. The results reported herein benefited from collaborations and/or information exchange within NASA's Nexus for Exoplanet System Science (NExSS) research coordination network sponsored by NASA's Science Mission Directorate, and includes data gathered with the 6.5 meter Magellan Telescopes located at Las Campanas Observatory, Chile. We acknowledge the contributions of Clare Leonard, Alex Watson, Elijah Spiro, Wyatt Mullen and Raymond Nzaba to the GAPlanetS pipeline. KMM's and LMC's work is supported by the NASA Exoplanets Research Program (XRP) by cooperative agreement NNX16AD44G. A.M. acknowledges the support of the DFG priority program SPP 1992 ``Exploring the Diversity of Extrasolar Planets" (MU 4172/1-1). L.M.C.'s research with MagAO was supported by the NSF ATI (Grant No. 1506818), the NSF AAG program \#1615408, and the NASA XRP program 80NSSC18K0441.






%


\begin{thebibliography}{}

\bibitem[Apai et al.(2016)]{Apai2016} Apai, D., Kasper, M., Skemer, A., et al.\ 2016, \apj, 820, 40 


\bibitem[Beuzit et al.(2008)]{Beuzit2008} Beuzit, J.-L., Feldt, M., Dohlen, K., et al.\ 2008, \procspie, 7014, 701418 


\bibitem[Bottom et al.(2017)]{Bottom2017} Bottom, M., Ruane, G., \& Mawet, D.\ 2017, Research Notes of the American Astronomical Society, 1, 30 

\bibitem[Bowler(2016)]{Bowler2016} Bowler, B.~P.\ 2016, \pasp, 128, 102001 

\bibitem[Close et al.(2008)]{Close2008} Close, L.~M., Gasho, V., Kopon, D., et al.\ 2008, \procspie, 7015, 70150Y 


\bibitem[Close et al.(2012)]{Close2012} Close, L.~M., Males, J.~R., Kopon, D.~A., et al.\ 2012a, \procspie, 8447, 84470X   

\bibitem[Close et al.(2013)]{Close2013} Close, L.~M., Males, J.~R., Morzinski, K., et al.\ 2013, \apj, 774, 94 


\bibitem[Close et al.(2014)]{Close2014} Close, L.~M., Follette, K.~B., Males, J.~R., et al.\ 2014, \apjl, 781, L30

\bibitem[Close et al.(2018)]{Close2018} Close, L.~M., et al. \ 2018, \procspie, 10703, https://arxiv.org/abs/1807.05070
 

\bibitem[D'Angelo et al.(2003)]{DAngelo2003} D'Angelo, G., Kley, W., \& Henning, T.\ 2003, \apj, 586, 540 

\bibitem[Dodson-Robinson \& Salyk(2011)]{Dodson2011} Dodson-Robinson, S.~E., \& Salyk, C.\ 2011, \apj, 738, 131 


\bibitem[Dong et al.(2012)]{Dong2012} Dong, R., Hashimoto, J., Rafikov, R., et al.\ 2012, \apj, 760, 111 

\bibitem[Eisner(2015)]{Eisner2015} Eisner, J.~A.\ 2015, \apjl, 803, L4 



\bibitem[Hashimoto et al.(2012)]{Hashimoto2012} Hashimoto, J., Dong, R., Kudo, T., et al.\ 2012, \apjl, 758, L19 

\bibitem[Henden et al.(2015)]{Henden2015} Henden, A.~A., Levine, S., Terrell, D., \& Welch, D.~L.\ 2015, American Astronomical Society Meeting Abstracts \#225, 225, 336.16 

\bibitem[Jayasinghe et al.(2018)]{asassn} Jayasinghe, T., Kochanek, C.~S., Stanek, K.~Z., et al.\ 2018, \mnras, 477, 3145 

\bibitem[Keppler et al.(2018)]{Keppler2018} Keppler, M., Benisty, M., M{\"u}ller, A., et al.\ 2018, arXiv:1806.11568 

\bibitem[Kley \& Nelson(2012)]{Kley2012} Kley, W., \& Nelson, R.~P.\ 2012, \araa, 50, 211 


\bibitem[Long et al.(2018)]{Long2018} Long, Z.~C., Akiyama, E., Sitko, M., et al.\ 2018, \apj, 858, 112 

\bibitem[Macintosh et al.(2006)]{Macintosh2006} Macintosh, B., Graham, J., Palmer, D., et al.\ 2006, \procspie, 6272, 62720L 


\bibitem[Males et al.(2014)]{Males2014} Males, J.~R., Close, L.~M., Morzinski, K.~M., et al.\ 2014, \apj, 786, 32 


\bibitem[Marois et al.(2006)]{Marois2006} Marois, C., Lafreni{\`e}re, D., Doyon, R., Macintosh, B., \& Nadeau, D.\ 2006, \apj, 641, 556 

\bibitem[Mawet et al.(2014)]{Mawet2014} Mawet, D., Milli, J., Wahhaj, Z., et al.\ 2014, \apj, 792, 97 


\bibitem[Meshkat et al.(2014)]{Meshkat2014} Meshkat, T., Kenworthy, M.~A., Quanz, S.~P., \& Amara, A.\ 2014, \apj, 780, 17 


\bibitem[Mordasini et al.(2017)]{Mordasini2017} Mordasini, C., Marleau, G.-D., \& Molli{\`e}re, P.\ 2017, \aap, 608, A72 


\bibitem[Morzinski et al.(2015)]{Morzinski2015} Morzinski, K.~M., Males, J.~R., Skemer, A.~J., et al.\ 2015, \apj, 815, 108 

\bibitem[Morzinski et al.(2016)]{Morzinski2016} Morzinski, K.~M., Close, L.~M., Males, J.~R., et al.\ 2016, \procspie, 9909, 990901 

\bibitem[M{\"u}ller et al.(2018)]{Muller2018} M{\"u}ller, A., Keppler, M., Henning, T., et al.\ 2018, arXiv:1806.11567 



\bibitem[Pecaut \& Mamajek(2016)]{Pecaut2016} Pecaut, M.~J., \& Mamajek, E.~E.\ 2016, \mnras, 461, 794 






\bibitem[Riaud et al.(2006)]{Riaud2006} Riaud, P., Mawet, D., Absil, O., et al.\ 2006, \aap, 458, 317 


\bibitem[Rigliaco et al.(2012)]{Rigliaco2012} Rigliaco, E., Natta, A., Testi, L., et al.\ 2012, \aap, 548, A56 

\bibitem[Sallum et al.(2015)]{Sallum2015} Sallum, S., Follette, K.~B., Eisner, J.~A., et al.\ 2015, \nat, 527, 342 





\bibitem[Soummer et al.(2012)]{Soummer2012} Soummer, R., Pueyo, L., \& Larkin, J.\ 2012, \apjl, 755, L28 

\bibitem[Uribe et al.(2013)]{Uribe2013} Uribe, A.~L., Klahr, H., \& Henning, T.\ 2013, \apj, 769, 97 




\bibitem[Wang et al.(2015)]{Wang2015} Wang, J.~J., Ruffio, J.-B., De Rosa, R.~J., et al.\ 2015, Astrophysics Source Code Library, ascl:1506.001 






\bibitem[Zhu(2015)]{Zhu2015} Zhu, Z.\ 2015, \apj, 799, 16 


\end{thebibliography}
\end{document}